\documentclass{article}

\usepackage{arxiv}
\usepackage{graphicx}
\usepackage{amsmath}
\usepackage{amssymb}
\usepackage{subfigure}
\usepackage{enumitem}
\usepackage{framed}
\usepackage{float}

\begin{document}

\title{A Closer Look at Mobile App Usage as a Persistent Biometric: A Small Case Study}

\author{Md A. Noor, G. Kaptan, V. Cherukupally, P. Gera, T. Neal\footnote{Cyber Identity and Behavior Research Lab Director and Assistant Professor, Tampa, FL 33620 USA. \textit{E-mail}: tjneal@usf.edu, December 2019} \\
Department of Computer Science and Engineering \\ 
University of South Florida}

\date{\nodate}

\maketitle

\begin{abstract}
In this paper, we explore mobile app use as a behavioral biometric identifier. While several efforts have also taken on this challenge, many have alluded to the inconsistency in human behavior, resulting in updating the biometric template frequently and periodically. Here, we represent app usage as simple images wherein each pixel value provides some information about the user's app usage. Then, we feed use these images to train a deep learning network (convolutional neural net) to classify the user's identity. Our contribution lies in the random order in which the images are fed to the classifier, thereby presenting novel evidence that there are some aspects of app usage that are indeed persistent. Our results yield a 96.8\% $F$-score without any updates to the template data. 
\end{abstract}

\section{Introduction} \label{introduction}
The number of smartphone users was estimated to exceed five billion at the time of this writing (2019) \cite{statistabillion}. Many use their devices for storing, transmitting, and manipulating sensitive data (e.g., bank records or work-related correspondences), resulting in several research efforts focusing on new and robust user authentication techniques for smartphone users. As a consequence, physical biometrics have grown increasingly popular due to commercialization in smartphones \cite{google, iphonefinger, samsungiris}. Behavioral, mobile biometrics have also received attention, particularly for authentication. Some of these efforts have explored the use of patterns found in mobile app usage as a behavioral biometric modality \cite{Li2014, qiao2016}. However, a literature review of these works (discussed in Section \ref{relatedwork}) suggests three assumptions: (1) data is constantly generated as a result of a users interaction with their device, (2) the most recent data is most reflective of the current user's usage patterns, and (3) the data is inherently non-permanent. 

Many \textit{continuous} recognition approaches are meant to counter the negative effects (e.g., low intra-person similarity) of these assumptions by maintaining a small temporal distance between template and query data to consequently account for change over time \cite{7503170, yang2015deep}. However, in this paper, we demonstrate an alternative technique to achieve satisfactory recognition performance despite such assumptions by exploring the possibility that persistent patterns of behavior actually \textit{do} exist. 

The idea of consistent patterns of behavior over an extended period of time can allow for innovative discoveries. By exposing permanent patterns of behavior, behavioral biometric modalities can be more useful for identification, and can then be used in multimodal, identification systems. For example, app usage may be integrated into face recognition systems used by law enforcement, and would be beneficial in cases where a phone is left at the crime scene with poor quality video surveillance.

We explore this hypothesis using the app data of a small set of smartphone users (15) as a preliminary case study of this approach. We first extract hand-engineered features from these data, and then convert our features into images. The pixels of the images encode time of day and the frequency of apps used, thereby representing multiple characteristics of behavior. The images are then fed into a convolutional neural network (CNN) in random order. Thus, the images themselves encode time, but the order in which the images are processed by the CNN is random. By training CNNs with images representative of hand-engineered features that are generated at random (in terms of time), we show that permanent patterns of behavior are learned by the network. To our knowledge, this is the first of such approach specific to observing the presence of persistent, biometric behaviors.

The paper is outlined as follows. Section \ref{relatedwork} discusses related work, Section \ref{data} describes the dataset used during experimentation, Section \ref{method} details our methodology, Section \ref{results} discusses the performance results and provides key insights regarding the use of behavioral imaging, and we conclude the paper in Section \ref{conclusion}.

\section{Related Work} \label{relatedwork}
Patterns found in phone usage have been extensively evaluated as behavioral biometric modalities since the inclusion of various sensors in smartphones. These sensors allow continuous data collection, such as touch, location, and app usage data, that have been found to be indicative of identity. For instance, Do and Gatica-Perez adapted stylometry-related bag-of-words and probabilistic topic modeling approaches to application traffic for identification of 111 subjects, each with device usage collected over one to eight months \cite{Do:2010:AYS:1899475.1899502}. The authors utilized voice, SMS, internet, camera, and gallery-related applications; 88.4\% precision was obtained when considering the true identity ranked among the top ten. 

With an increased use of artificial neural networks, additional analyses employing these networks have emerged. For instance, an autoencoder was employed as a signal extraction tool for mobile device users by Rejashekar et al. \cite{7836683}. The encoding was then fed into a self-organizing map (SOM) in which behavior was further characterized. This methodology was tested on the LiveLab database consisting of iPhone traces from 25 subjects. The most frequented applications, websites, and cell towers were used as input into the two-network framework; the dissimilarity between two subjects was determined by the chi-squared distance between the outputs of a one-class trained SOM. An average dissimilarity percentage of 74\% was reported, with the best recognition capability reflected as 86\% dissimilarity. 

As discussed in Section \ref{introduction}, many mobile biometric systems consider continuous recognition approaches, where new, incoming data is used to maintain updated templates. For example, Yao et al. \cite{yao2017continuous} developed an adaptive neuro-fuzzy inference system (ANFIS) for continuous authentication. Features included screen status, incoming and outgoing calls, Wi-Fi and browsing history, incoming and outgoing SMS, and app history. While this work leveraged the power of deep learning by merging fuzzy logic and neural networks to achieve 95\% accuracy, a continuous recognition framework processes data over time, and does not attempt to identify the permanent information in the data stream.

Finally, several works have explored the use of mobile apps in unimodal biometric systems. Mahbub et al. \cite{mahbub2018continuous} explored state space models on app usage  \cite{mahbub2018continuous, mahbub2016active}, and a ranking method developed by Alzubaidi et al. \cite{alzubaidi2018data} was derived based on the informativeness and popularity of mobile apps. In addition, Canfora et al. \cite{canfora2017mobile} used app usage from 15 users to develop transition signatures from one app to another as biometric markers. Thus, patterns in app usage is not a novel biometric modality, but stable signatures existing within these patterns have yet to be realized. 

\section{Data} \label{data}

\begin{table*}[!ht]
\centering
\caption{50 most frequented mobile apps across all 15 users. The third and sixth columns show the corresponding normalized frequencies for each mobile app.}
\label{tbl:globalFreqs_top50}
\scalebox{0.6}{
{\renewcommand{\arraystretch}{1.2}%
\begin{tabular}{|l|l|l|l|l|l|}
\hline
1 & com.google.process.gapps & 0.091300257 & 26 & com.weather.Weather & 0.008645851 \\ \hline
2 & com.netscale.phonemonitor & 0.081385704 & 27 & com.pandora.android & 0.00826731 \\ \hline
3 & com.facebook.katana & 0.070359124 & 28 & com.lookout & 0.008166526 \\ \hline
4 & com.google.android.gm & 0.069263895 & 29 & com.droid27.senseflipclockweather & 0.006385804 \\ \hline
5 & com.android.phone & 0.066118785 & 30 & com.usatoday.android.news:com.urbanairship.push.process & 0.006230728 \\ \hline
6 & com.google.android.gsf.login & 0.05403887 & 31 & com.whatsapp & 0.006004963 \\ \hline
7 & com.google.android.apps.googlevoice & 0.036389223 & 32 & com.zynga.scramble & 0.005759502 \\ \hline
8 & com.google.android.apps.maps & 0.03320792 & 33 & com.locationlabs.v3client & 0.005627022 \\ \hline
9 & com.cyanogenmod.stats & 0.029236039 & 34 & com.coremobility.app.vnotes & 0.005366763 \\ \hline
10 & android.process.media & 0.024799131 & 35 & com.htc.launcher & 0.005080807 \\ \hline
11 & com.android.browser & 0.024747339 & 36 & com.redbend.vdmc & 0.005018417 \\ \hline
12 & com.google.android.apps.maps:GoogleLocationService & 0.023635713 & 37 & com.espn.espnfantasyfootball:com.urbanairship.push.process & 0.004457205 \\ \hline
13 & com.android.vending & 0.019445866 & 38 & com.htc.weather.bg & 0.004285831 \\ \hline
14 & com.twitter.android & 0.015616164 & 39 & com.google.android.apps.maps:NetworkLocationServic & 0.004283232 \\ \hline
15 & com.google.android.apps.maps:LocationFriendService & 0.014920672 & 40 & com.facebook.katana:providers & 0.004259635 \\ \hline
16 & com.google.android.apps.maps:FriendService & 0.014787193 & 41 & com.netscale.quiz & 0.004213343 \\ \hline
17 & com.espn.score\_center & 0.013336719 & 42 & com.usatoday.android.news & 0.004150553 \\ \hline
18 & com.facebook.orca & 0.012890089 & 43 & com.jb.gosms & 0.00404127 \\ \hline
19 & com.google.process.location & 0.012723415 & 44 & twc.weatherController & 0.003795508 \\ \hline
20 & com.google.android.gms & 0.011936538 & 45 & com.google.android.gallery3d & 0.003556545 \\ \hline
21 & com.android.mms & 0.011085671 & 46 & com.cyanogenmod.cmparts & 0.003535249 \\ \hline
22 & com.google.android.apps.maps:NetworkLocationService & 0.011058875 & 47 & tunein.service & 0.003455761 \\ \hline
23 & com.google.android.gcm & 0.010913098 & 48 & com.google.android.calendar & 0.003376073 \\ \hline
24 & com.google.android.youtube & 0.010603246 & 49 & com.groupme.android & 0.003290987 \\ \hline
25 & com.google.android.apps.genie.geniewidget & 0.010197209 & 50 & com.android.email & 0.003199701 \\ \hline
\end{tabular}}}
\end{table*}

The data used in this paper contains logs of mobile app use from 15 students using Android devices. The data was provided by colleagues from another university; as a consequence, we are limited in the details regarding data collection (e.g., brands of devices used). Thus, this section provides the statistics regarding data gathered from the individual logs of mobile app usage. Each log entry contains the package name for an app event with the associated date and time in which the app was either receiving or transmitting data. The minimum and maximum number of app events across all logs are 196,462 and 1,137,711, respectively, with an average of 666,771 events. Starting dates varied for data collection across the 15 users from January 2, 2012 to May 5, 2012. End dates also varied from November 1, 2012 to September 1, 2013. The average number of days recorded per user is 297 with a standard deviation of 70. The smallest log in terms of days is 168, and the largest consists of 414 days. There are 443 total unique apps across all 15 users. The minimum and maximum number of unique apps per user are 38 and 138, respectively, with an average of 81. Table \ref{tbl:globalFreqs_top50} shows the top 50 apps across all users and their normalized frequencies.

\section{Apps to Images} \label{method}
We started our experiments by constructing images from each user's mobile app usage. To do this, we first converted every app (using the app's packing name) into a numerical value using \texttt{scikit-learn}'s LabelEncoder\footnote{https://scikit-learn.org/stable/modules/generated/sklearn.preprocessing.LabelEncoder.html}. Then, we generated one image per day (12:00AM through 11:59PM) for each user. The placement of the pixels along an image's $x$-axis correspond with the numerically-encoded apps (i.e., pixels in column 10 give information about app 10). The $y$-axis represents time in minutes, meaning that the pixel intensity at location ($i$, $j$) provides information for app $i$ accessed by the user at time $j$. 

To prevent elongated, rectangular images ($443\times1440$), we considered app activity every three minutes (e.g., an app event between 12:00 and 12:03AM with an encoded value of 12 corresponds with the pixel at ($12, 1$)). Images were $443\times480$ dimensions as a result. The color of a pixel at ($i$, $j$) depended on how often the user accessed app $i$ during the time span at $j$. We considered three approaches for measuring frequency, producing three disparate sets of images: \texttt{GLOBAL}, \texttt{LOCAL}, and \texttt{PERDAY}. 
\begin{itemize}[noitemsep, topsep=0pt]
    \item \texttt{GLOBAL}: The frequency to all 443 apps across the entire dataset of 15 users was computed, and each app's normalized frequency was used as its pixel value. Given app frequencies $f_1 ,f_2, f_3, \ldots, f_{443}$, the normalized frequency, $f_{a_i}$ of app $a$ with a frequency of $f_a$ was computed as     \[ \frac {f_a}{\sum_{i=1}^{443} f_i} \].
    
    \item \texttt{LOCAL}: The normalized frequency to all 443 apps per user was used as the app's pixel value. 
    
    \item \texttt{PERDAY}: The normalized frequency to all 443 apps per user per day was used as the app's pixel value. Note that the pixel value for an app for a single user may change day to day.
\end{itemize}

\begin{figure*}[ht]
\centering
\subfigure[\texttt{GLOBAL}, User A, Day 100]{
\includegraphics[width=0.3\columnwidth]{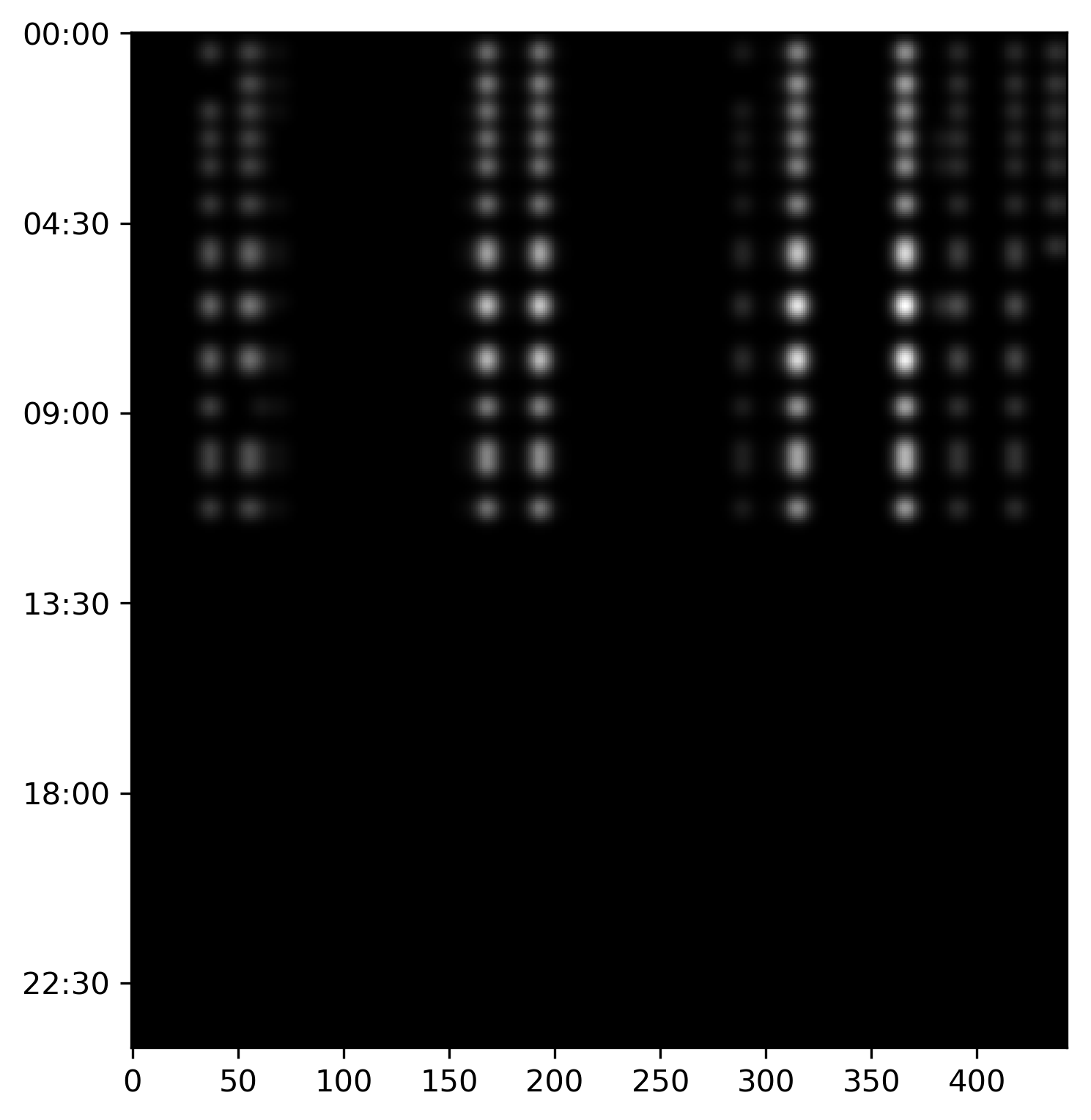}
\label{global}
}
~
\subfigure[\texttt{LOCAL}, User A, Day 100]{
\includegraphics[width=0.3\columnwidth]{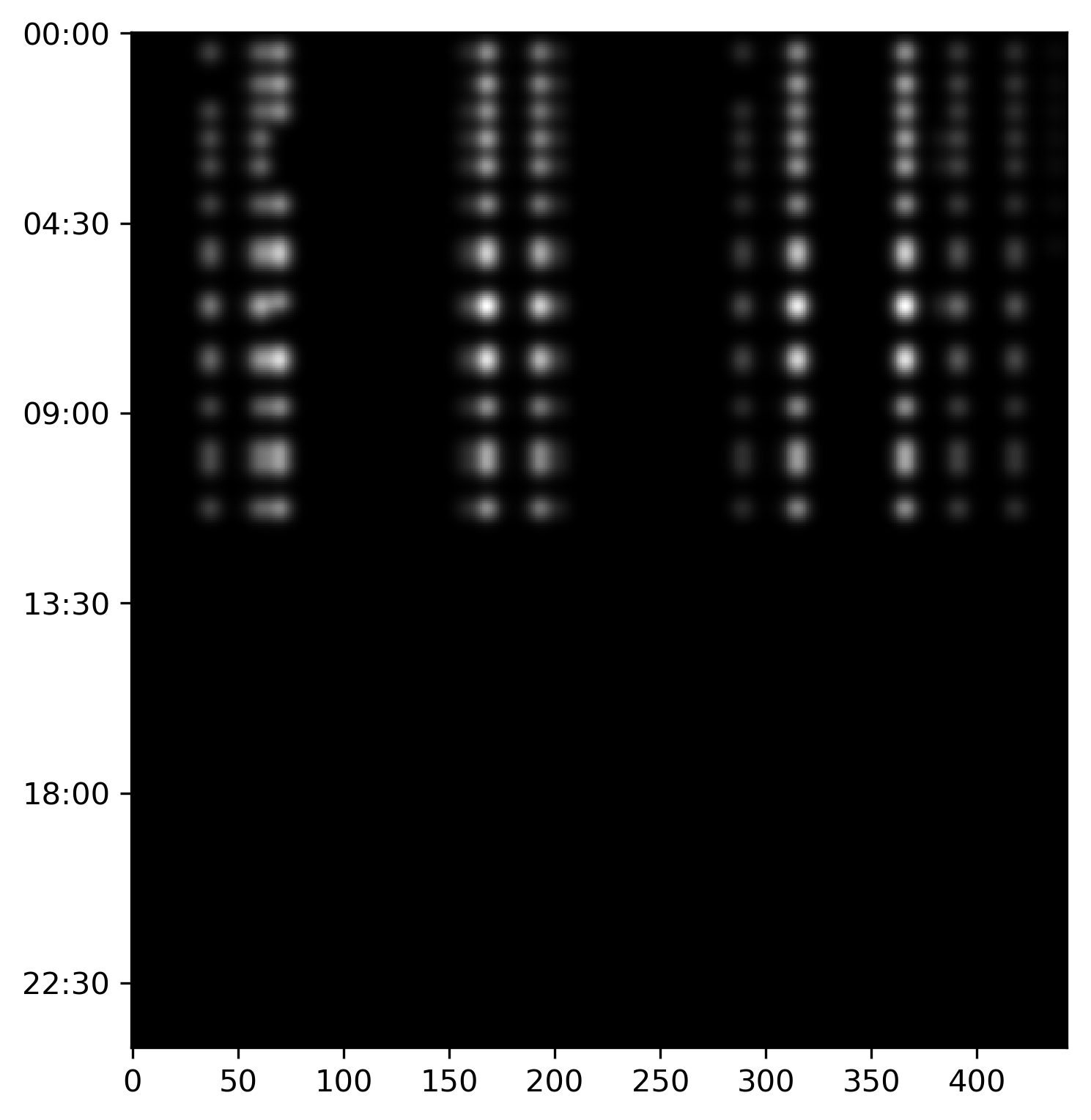}
\label{local}
}
\subfigure[\texttt{PERDAY}, User A, Day 100]{
\includegraphics[width=0.3\columnwidth]{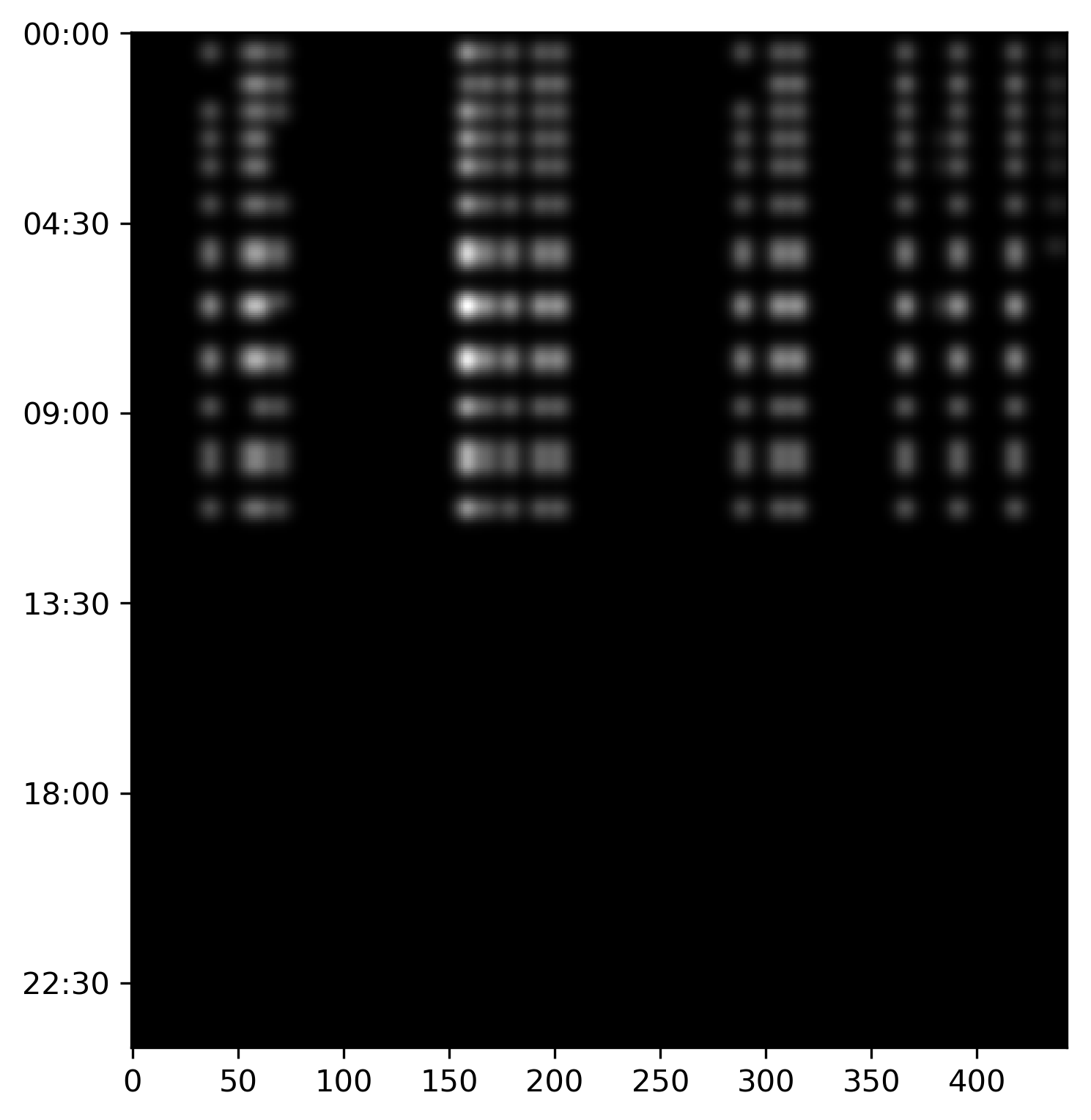}
\label{perdayFreq}
}
\caption{Subfigures \ref{global}, \ref{local}, and \ref{perdayFreq} shows the \texttt{GLOBAL}, \texttt{LOCAL}, and \texttt{PERDAY} images for the same user in a single day. These images show that as the calculation for pixel intensity changes, the images differ, although they represent the same data for a single user.} 
\label{img:experiment_examples}
\end{figure*}

Obviously, a user may be quite active within a three minute period, visiting more than a single app. When this occurs, we considered the frequencies of each unique app event within the three-minute span. For instance, when generating a \texttt{GLOBAL} image, assume that there are five app events, $a_1, a_2, \ldots, a_5$, for a particular three-minute window. Their frequencies within the three-minute span are $f_1, f_2, \ldots, f_5$, and \texttt{GLOBAL} frequencies are $f_{g_1}, f_{g_2}, \ldots, f_{g_5}$. The final frequency, or pixel intensities, $f_{a_i}$, of an application $a_i$ occurring within that three-minute span is computed using Equation \ref{eq:final_freq_eq}. Figure \ref{img:experiment_examples} provides example images.

\begin{equation} 
\label{eq:final_freq_eq}
    f_{a_i} = \frac {f_i * f_{g_i}}{\sum_{i=1}^{5} f_i * f_{g_i}} 
\end{equation}

\begin{figure*}[ht]
\centering
\subfigure[Filter Size 1]{
\includegraphics[width=0.3\columnwidth]{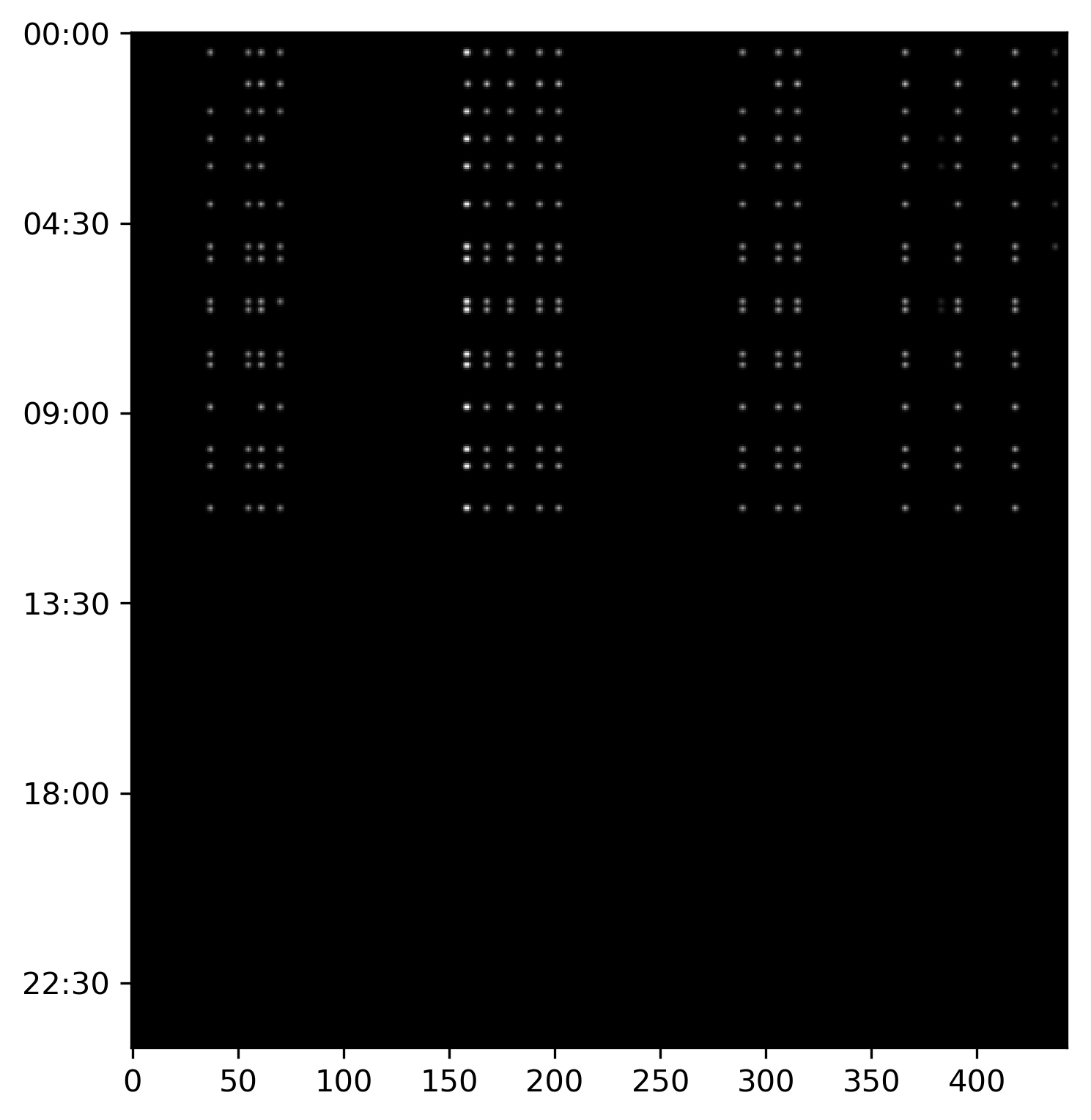}
\label{filter1}
}
\subfigure[Filter Size 8]{
\includegraphics[width=0.3\columnwidth]{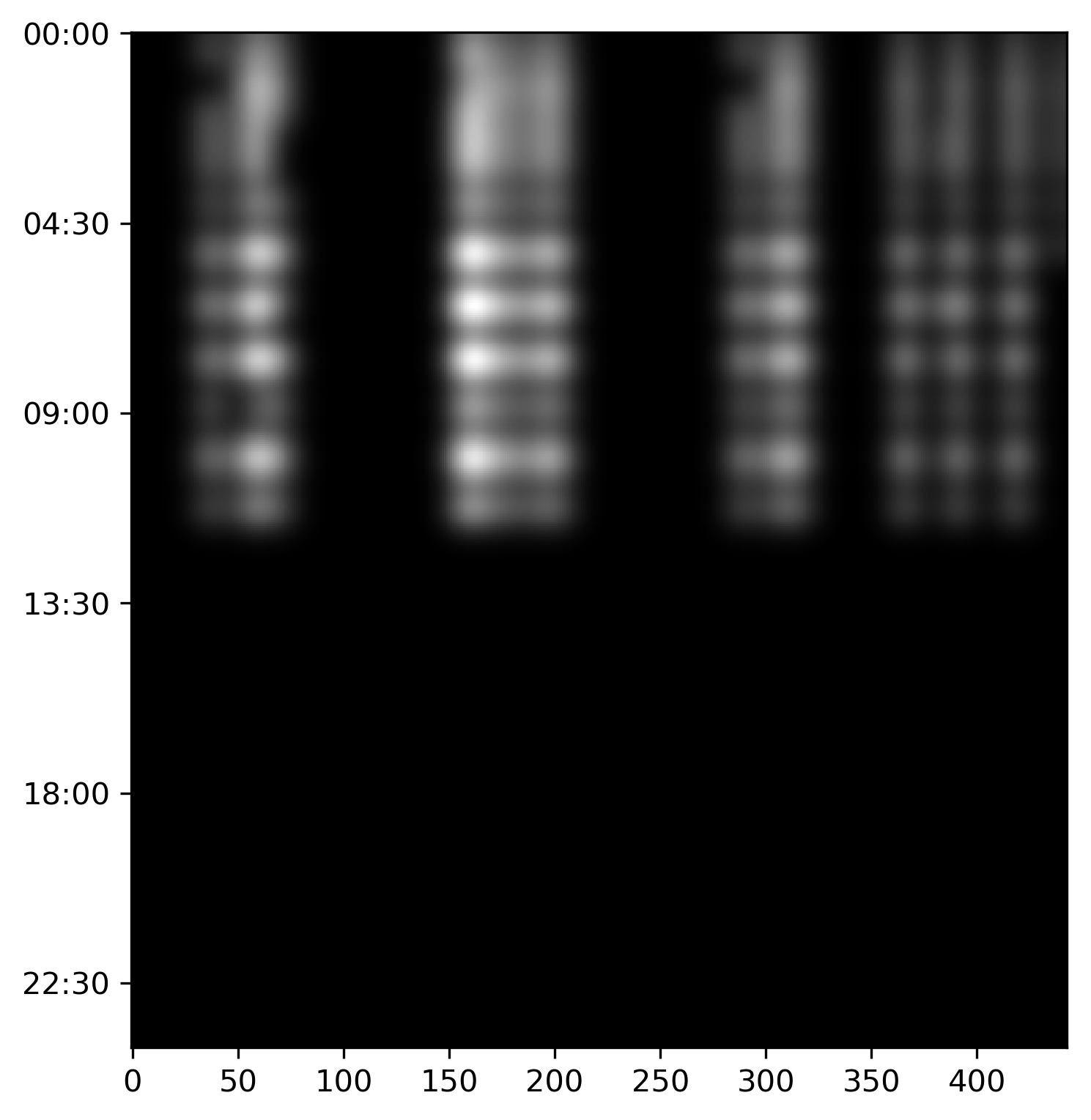}
\label{filter8}
}
\subfigure[Filter Size 15]{
\includegraphics[width=0.3\columnwidth]{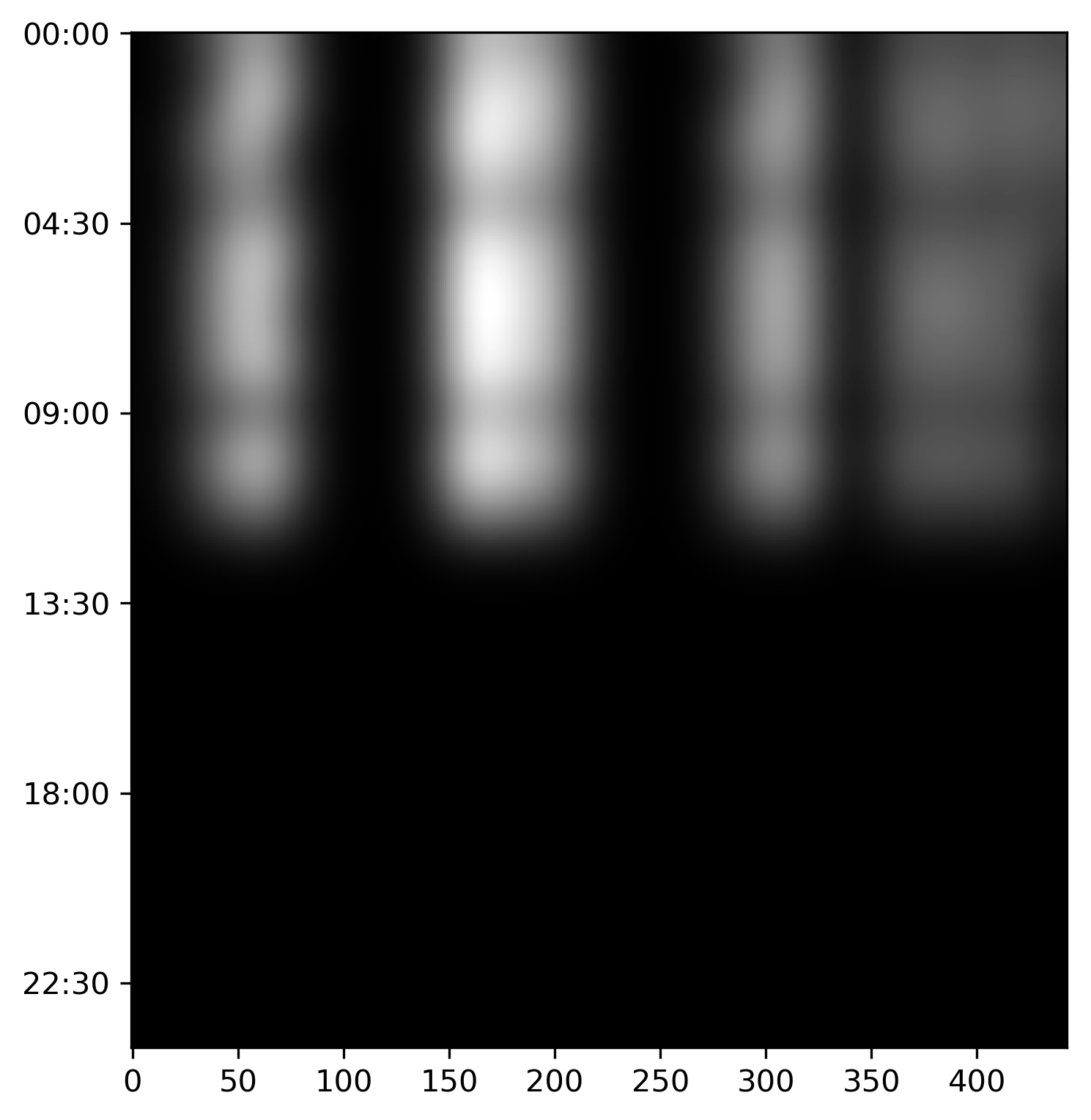}
\label{filter15}
}

\caption{Gaussian filtered images for User A on the same day.} 
\label{img:filter_size}
\end{figure*}

The aforementioned process resulted in three sets of images, each consisting of 4,451 images. Since deep neural networks typically require several training samples per class, we used Gaussian filters of various sizes to generate additional images. For every image initially produced, we applied 15 Gaussian blur filters of sizes 1 through 15. This particular filtering approach was chosen not only to increase the number of training samples, but to reduce the information regarding exact app activity within the images. As filter sizes grew larger, the smoothing produced by the filters minimized the ability to visually recognize the exact apps that were used by the user. The Gaussian filters instead blurred sharp edges in the images, leaving time-specific information about the image (i.e., time of day when the user is most/least active). Figure \ref{img:filter_size} shows images of three different filter sizes applied in a particular day. Filtering produced three final sets of images, each consisting of 66,765 images. This is a sufficient amount of data for training neural networks, while each image offers unique information regarding the user's activity patterns.

\subsection{Training the CNN}
Once the images were prepared, we resized each image to $50\times50$ dimensions, and then shuffled and split the images into five folds for 5-fold cross-validation. Shuffling the data prior to splitting into 5-folds was meant to impart randomness, such that any fold would not contain all images produced from consecutive days which are more likely to consist of similarities in the placement of bright-colored pixels. Further, our choice of CNNs over networks that can learn sequences (e.g., Recurrent and Long-Short Term Memory neural networks) is to further reduce the network's ability to learn app usage patterns that are dependent on what has previously occurred. Hammerla et al. \cite{hammerla2016deep} also argued that ``recurrent networks outperform convolutional networks significantly on activities that are short
in duration but have a natural ordering, where a recurrent
approach benefits from the ability to contextualise observations across long periods of time. For prolonged and
repetitive activities... we recommend to
use CNNs.'' Since our aim is to uncover prolonged activities, we have pursued the analysis of CNNs to explore the hypothesis that persistent app usage patterns can exist. We used 20\% of each training fold as validation data. The CNN architecture is provided in Figure \ref{cnn_config}. Our network consists of three convolution layers, utilizing max pooling and a Leaky ReLU activation function. 

Finally, we conducted four variations of experiments using the CNN architecture for each of the three sets of images (i.e., \texttt{GLOBAL}, \texttt{LOCAL}, and \texttt{PERDAY}) for a total of 12 experiments:
\begin{enumerate}[noitemsep, topsep=0pt]
\item \texttt{ALL}: These experiments involved all images. \item \texttt{F1-F7}: These experiments only used images produced with Gaussian filter sizes one through seven. Since these images have less blur, they provide more information regarding which apps were being used.
\item \texttt{F8-F15}: These experiments only used images produced with Gaussian filter sizes eight through 15. Since these images were significantly affected by smoothing, it is more difficult to visually determine which apps were used. However, the time of day in which users were active is obvious.
\item \texttt{DROPOUT}: To simulate the scenario where users are inactive, we randomly removed 20\% of the images, and then split the data into five folds. 
\end{enumerate}

\begin{figure}
\centering
\footnotesize{
\begin{verbatim}
Layer                   Output Shape      
==========================================
Conv, 32 3x3 kernels    (50, 50, 32)           
__________________________________________
LeakyReLU, alpha=0.1    (50, 50, 32)             
__________________________________________
MaxPool, 2x2            (25, 25, 32)             
__________________________________________
Dropout, 0.20           (25, 25, 32)           
__________________________________________
Conv, 32 3x3 kernels    (25, 25, 32)          
__________________________________________
LeakyReLU, alpha=0.1    (25, 25, 32)           
__________________________________________
MaxPool, 2x2            (12, 12, 32)             
__________________________________________
Dropout, 0.30           (12, 12, 32)             
__________________________________________
Conv, 64 3x3 kernels    (12, 12, 64)         
__________________________________________
LeakyReLU, alpha=0.1    (12, 12, 64)            
__________________________________________
MaxPool, 2x2            (6, 6, 64)               
__________________________________________
Dropout, 0.30           (6, 6, 64)               
__________________________________________
Flatten                 (2304)                  
__________________________________________
Dense, ReLU             (512)              
__________________________________________
Dropout, 0.50           (512)                  
__________________________________________
Dense, Softmax          (15)                  
==========================================
Trainable params: 1,215,919
Non-trainable params: 0
Batch size / epochs / stride: 64 / 20 / 1
Optimizer: RMSprop
Loss Function: Categorical cross-entropy
\end{verbatim}
}
\caption{Convolutional neural network configuration with three convolution layers.}
\label{cnn_config}
\end{figure}

\section{Results} \label{results}
We evaluated identification performance using the $F$-score, the harmonic mean of precision and recall. If $TP$, $FP$, and $FN$ correspond with a true positive, false positive, and false negative, respectively, then the $F$-score, precision, and recall are defined as follows:
\[ F-score = 2 \times \frac{precision \times recall}{precision + recall} \]
\[ Precision = \frac{TP}{TP + FP} \]
\[ Recall = \frac{TP}{TP + FN} \]

\begin{figure}[!ht]
\centering
\includegraphics[width=0.6\columnwidth]{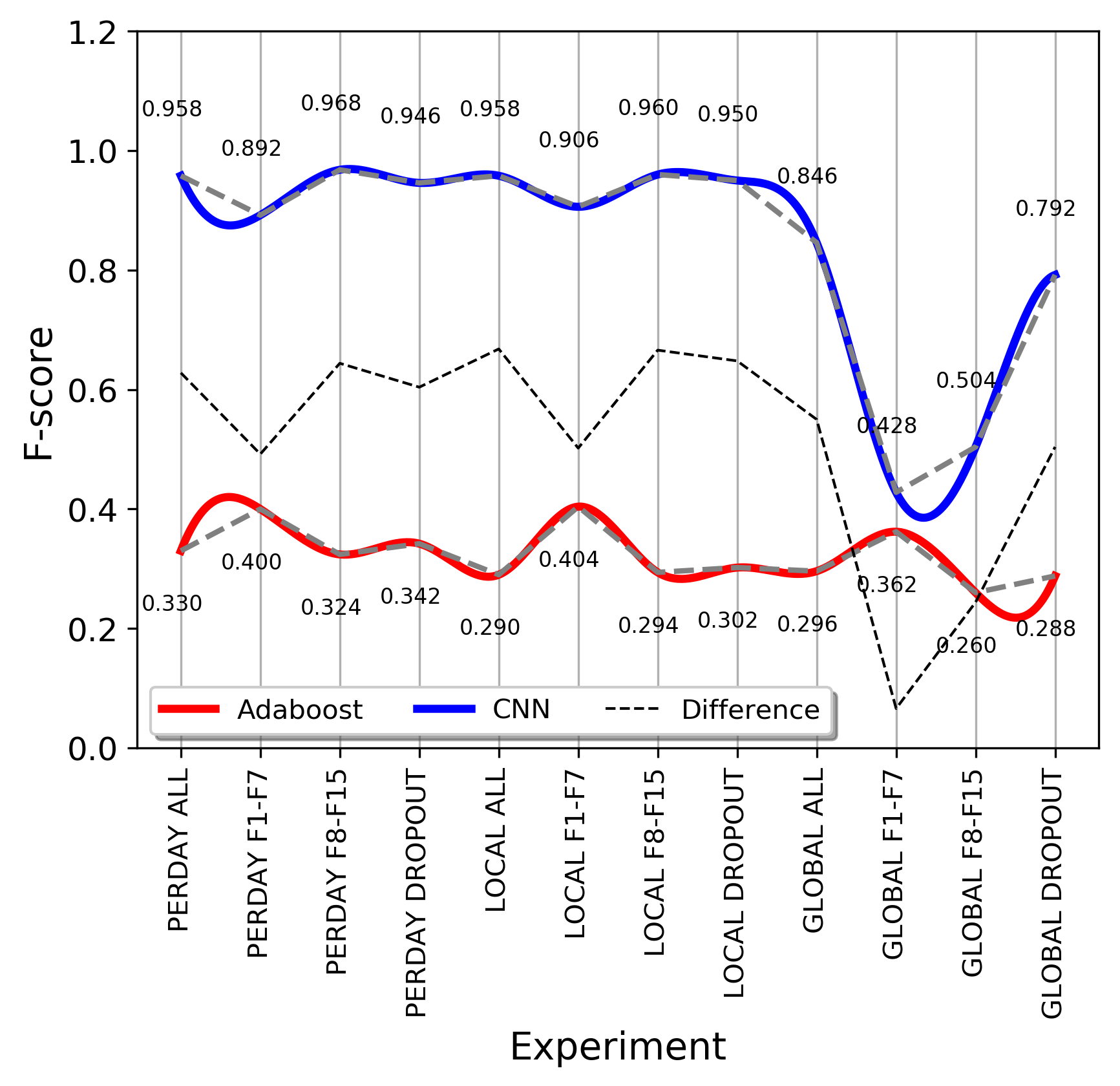}
\caption{$F$-scores using a convolutional neural network and Adaboost classifiers.} 
\label{img:performance}
\end{figure}

As a baseline, we compared the use of a CNN with an Adaboost\footnote{https://scikit-learn.org/stable/modules/generated/sklearn.ensemble.AdaBoostClassifier.html} classifier using the default parameters in \texttt{scikit-learn}. The input to the Adaboost classifier was the same as the CNN, except the images (pixel values) were converted to 1D vectors. There were three key differences in performance associated with the use of images and deep neural networks compared to relying upon hand-engineered features alone drawn from the results presented in Figure \ref{img:performance} that are discussed in this section. 

\subsection{Timing Versus Choice}
Our findings suggest that while hand-engineered features provide information regarding the user's choice of apps, CNNs learn features that are more reflective of timing. Figure \ref{img:performance} shows a decrease in $F$-scores from \texttt{F1-F7} to \texttt{F8-F15} using \texttt{PERDAY}, \texttt{LOCAL}, and \texttt{GLOBAL} images for Adaboost (i.e., hand-engineered features only). However, there is an increase in $F$-scores when using the CNN. Recall that experiments \texttt{F1-F7} utilized images generated with Gaussian filters sized one to seven. These images have less blur, and are therefore more indicative of which app was used according to the location of the pixel along the $x$-axis. The smaller the filter size, the less each pixel intensity spreads to neighboring pixels; thus, it is visually easier to recognize specific apps that were used. 

\begin{figure*}[!htp]
\centering
\subfigure[Test image 1]{
\includegraphics[width=0.45\columnwidth]{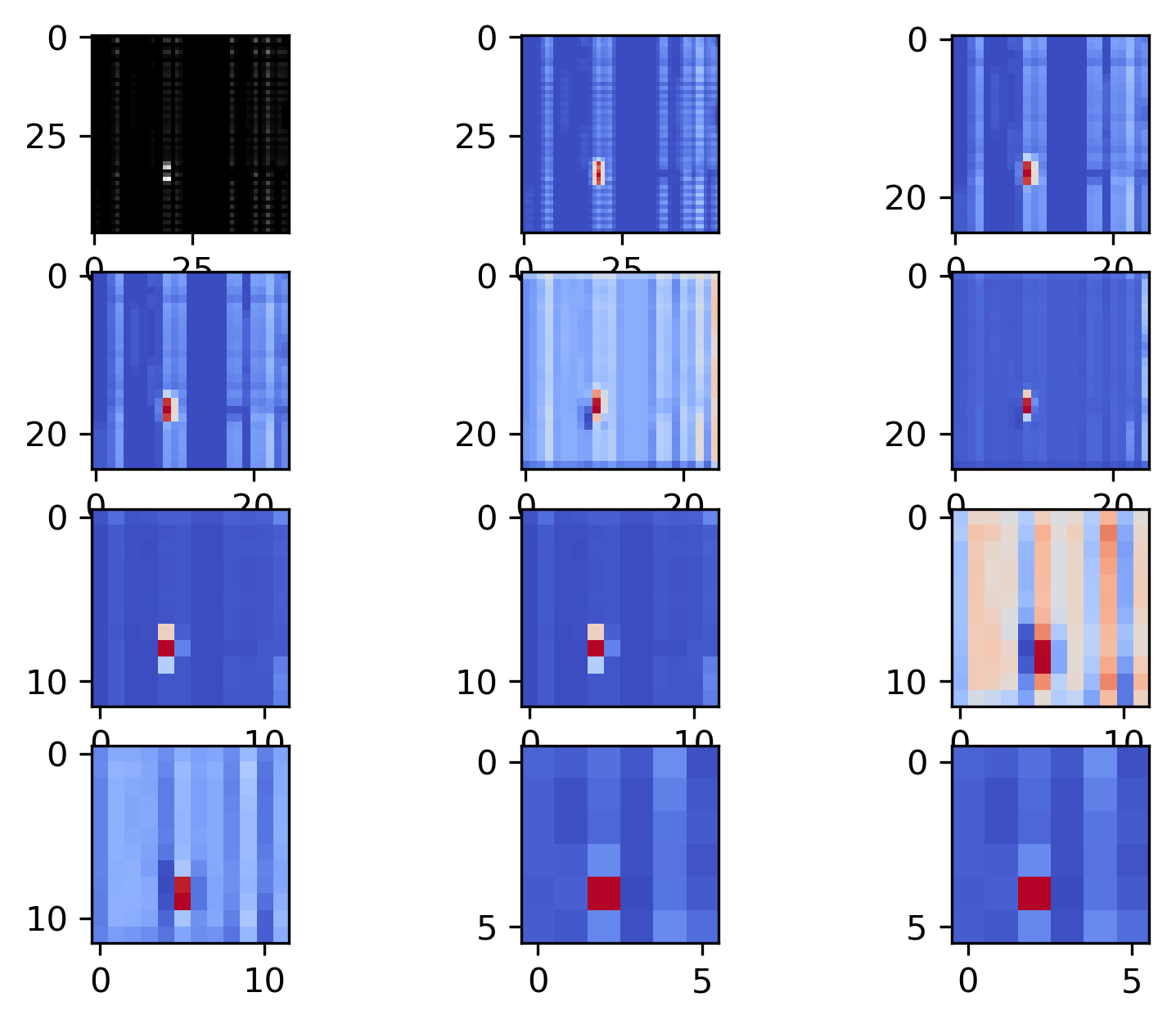}}
\subfigure[Test image 2]{
\includegraphics[width=0.45\columnwidth]{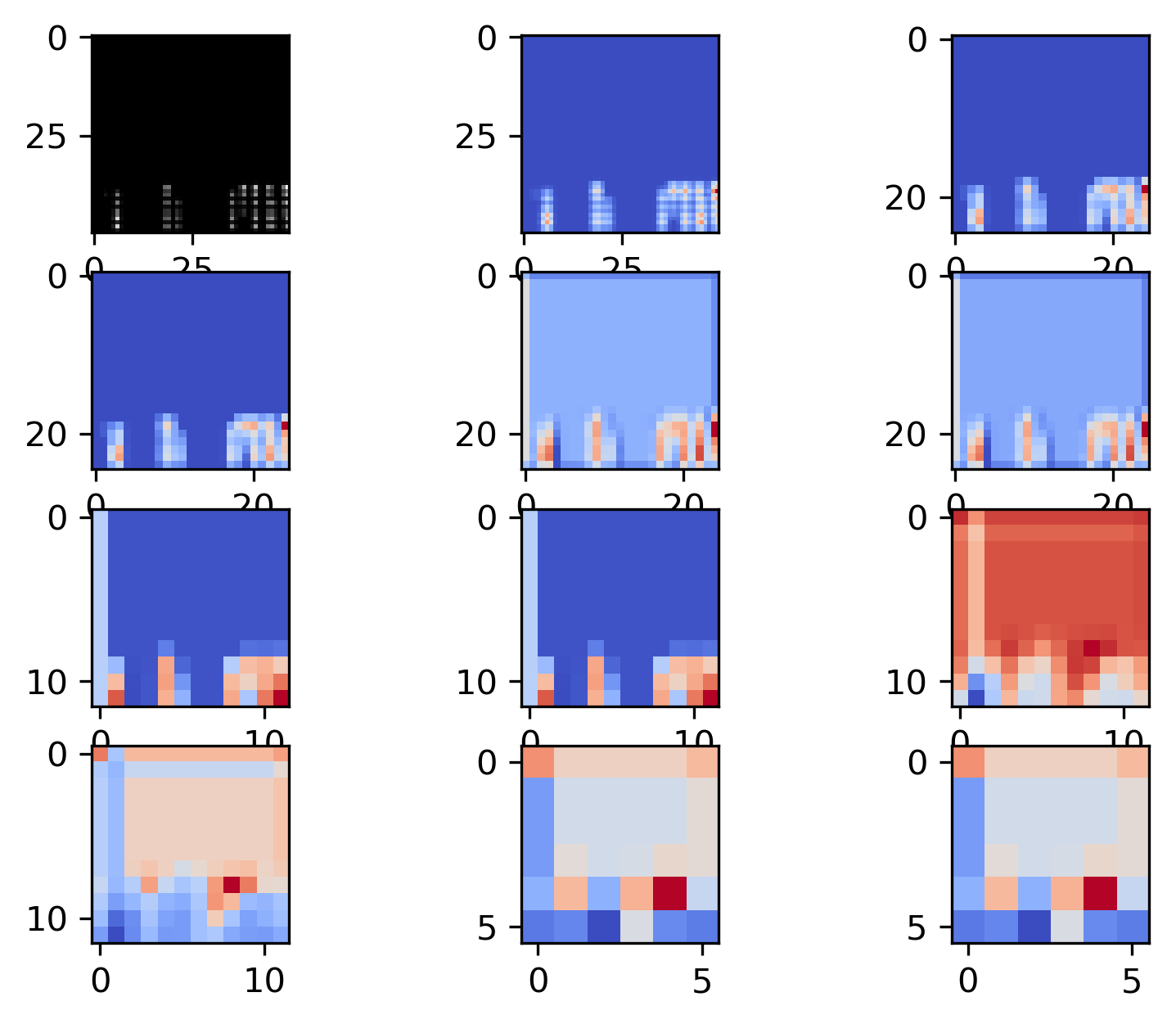}}
\caption{CNN outputs for two test images. Although it is visually possible to determine which apps are used by mapping the pixel values to their location on the $x$-axis, the network incrementally generalizes the user's activity into an approximation of which app is used with an overall depiction of when the user was active.} 
\label{img:timingVSchoice}
\end{figure*}

On the other hand, for larger filter sizes (i.e., \texttt{F8-F15}), the amount of blur per pixel spreads across a larger area such that if multiple apps with a similar numerical encoding were used around the same time of day, it is visually more difficult to determine which and how many apps were used. We do know that the wider the blur, the more apps that were used by the user. Further, the brighter the blur, the more likely that the apps that were used were frequently used across the dataset (i.e., \texttt{GLOBAL}) and/or by that user (i.e., \texttt{LOCAL} or \texttt{PERDAY}). Thus, experiment \texttt{F1-F7} can be interpreted as providing information regarding app usage, while experiment \texttt{F8-F15} can be thought of as providing information regarding time of app usage. This is observed in Figure \ref{img:timingVSchoice} which depicts the layers of the network for two test images taken from \texttt{PERDAY F1-F7}; the pixel values in the input image (upper left) essentially correlate with exact app usage and frequency (from pixel intensity), but throughout the network, the layer outputs grow into patch-like representations more suggestive of when the user was active. 

Since the image-based neural network approach outperforms the use of hand-engineered features alone, this may be an indication that permanent behaviors may be linked to activity levels across the day. This finding provides significant insight to the problem of behavioral biometrics; although we have restricted our analysis to mobile app usage, our findings show that people may possess a pattern of behavior that does \textit{not} significantly deviate over time.

\subsection{Hand-Engineered Features Matter}
\texttt{GLOBAL}, \texttt{LOCAL}, and \texttt{PERDAY} images differ by the pixel intensities for the same app. For \texttt{GLOBAL} experiments, the pixel intensity for each app was computed according to the frequency of each app across the entire dataset. The pixel intensity per app was computed per user for \texttt{LOCAL} experiments, and the pixel intensity per app was computed daily per user for \texttt{PERDAY} experiments. In Figure \ref{img:performance}, we find that performance significantly worsens for the global representation when using a CNN. This is also the point at which the difference between scores for the CNN and Adaboost classifiers minimizes (denoted by the dotted black line). 

Since the ``placement'' of the app does not change between image types, we attribute this observation to the changes in pixel intensities that correlate with intra- and inter-person similarity. \texttt{PERDAY} and \texttt{LOCAL} experiments encode pixel values according to intra-person app usage. Given our results, we argue that these two representations help to minimize inter-person similarity while providing detailed information regarding subjects' use of mobile apps. However, the global representation contributes to inter-person similarity given the overall encoding of pixel values. To optimize performance using the global representation, two different subjects that use the same apps would have to use these apps at different times in the day to be distinguishable. 

Our findings from \texttt{GLOBAL} images are an indication that many users may use the same apps around the same time of day. Since the CNN learned ``patches'' of activity, or times of day \textit{when} a user is active with an approximation (due to blurring) of \textit{what} the user is doing, if the numerical encodings (which we achieved using \texttt{scikit-learn}'s LabelEncoder) for several apps place them in similar locations within the image along the $x$-axis, the intensity of the ``patches'' for one user will then resemble the patch of others considering the shared pixel intensities in \texttt{GLOBAL} images. Therefore, hand-engineered features provide important information into the network. The domain-specific knowledge of user activity is leveraged by the CNN as demonstrated by changes in performance across the three image types.

\subsection{The Impact of Inactivity}
In this discussion, we highlight the impact of \texttt{DROPOUT} on identification performance. These experiments were intended to replicate scenarios where users were inactive, thereby limiting the amount of data available for recognition. Here, performance worsens for both the CNN and Adaboost classifiers, but by different degrees. In the CNN, $F$-scores decreased by an average of 3.47\%; there was a 7.8\% decrease in $F$-scores when using Adaboost. We computed these differences using the best $F$-score achieved per image type versus the $F$-score for \texttt{DROPOUT}. Thus, our findings demonstrate a significant advantage associated with a CNN in situations where data is unavailable. Our intent here is to demonstrate that even when users are inactive, a thread of behavior that remains consistent over time can be learned. This is a valuable insight, especially considering that many continuous recognition schemes incorporate some methodical measure to cope with missing data (e.g., \cite{7139043, 7791164}).

\section{Conclusion} \label{conclusion}
In this paper, we explored the mobile app use from a small set of users (15) for identifying persistent patterns of app use that could be of value for biometric systems. We represented the app data as images; each image represented domain knowledge of a user's activity levels. Our results show that CNNs can learn persistent patterns of behavior, but mostly focus on \textit{when} apps were used compared to \textit{what} apps were used, with a 96.8\% $F$-score. 

Expansions of this work reflect some of the observed limitations of our experiments. Although the achieved performance indicates that CNNs are capable of learning persistent patterns of behavior, we did not discuss these behaviors in detail. Our analysis suggests that persistent patterns of behavior are correlated with activity level across the day (e.g., a person is generally active at night); future work should attempt to derive a methodology that can define these patterns in more detail. Second, we compare a CNN to an Adaboost classifier; while the intent is to demonstrate that learned features are useful for identifying persistent behaviors, our approach requires additional comparisons against more classifiers for a more complete validation. Third, based on the results presented in this paper, we hypothesize that there are multiple persistent patterns of behavior. We limited this work to frequencies across the day; in future work, we will consider additional hand-engineered features.

\bibliographystyle{plain}
\bibliography{refs}

\end{document}